# "Pattern of Texture Zeroes in Quark Mass Matrices With a Divergent Top-Yukawa Coupling at the GUT Scale"

Bipin R. Desai[1] and D. P. Roy[1,2]

[1] *Department of Physics, University of California*
*Riverside, California 92521, USA*

[2] *Tata Institute of Fundamental Research, Mumbai 400 005, India*

**Abstract**

In a SUSY GUT model responsible for generating symmetric quark mass matrices at the GUT scale ($\mu = \wedge$) we assume that the top-Yukawa coupling, $\lambda_t$, becomes infinite at that scale. As a consequence, the MSSM renormalization group equations for quark Yukawa couplings exhibit hierarchical solutions which lead to a pattern of texture zeroes in quark mass matrices at $\mu = \wedge$ similar to one of the solutions of Ramond, Roberts and Ross. The evolution in energy scale to low energies shows excellent agreement between the measured quantities involving the scale-independent ratios of CKM matrix elements and their predicted values in terms of quark mass ratios. It is, noted that the t condensate model of Bardeen, et al., predicts an infinite $\lambda_t$ at $\mu = \wedge$ implying, for our model, that at $\mu = \wedge$, both the symmetry of Yukawa matrices and condensate dynamics may have a common origin.

# I. INTRODUCTION

In the quark-sector of the Standard Model, to which we will confine our attention, there are ten parameters that are arbitrary but experimentally measurable: six quark masses, along with three mixing angles and one phase angle of the CKM (Cabbibo-Kobayashi-Masakawa) matrix. An enormous body of literature exists today dealing with attempts to reduce this arbitrariness by invoking simplified structures for the up (U) and down (D) - sector Yukawa matrices at a higher scale (e.g., the Grand Unified Theory (GUT) scale). [1][2]

Typically, the 3 x 3 Yukawa matrices are assumed to be symmetric at the GUT scale and to have zeroes at specific locations ("texture zeroes")[3]. While the symmetry of the Yukawa matrices is a natural assumption in Grand Unified Theories at the level of SO(10) and beyond, there is as yet no obvious mechanism that can accommodate the texture zeroes.

The above two assumptions, however, reduce the number of independent parameters allowing relations to develop at the GUT scale between quark masses, which are the eigenvalues of the Yukawa matrices, and the mixing angles, which are related to the rotation matrices diagonalizing these Yukawa matrices. These relations are then evolved, down to low-energy scales, using the renormalization group (RG) equations, and compared with experiments.

One such model, which we will frequently refer to, is that of Ramond, Roberts, and Ross (RRR) [2] in which the structures of U and D are systematically analyzed with maximum number of allowed texture zeroes. They find that there are five possible solutions at the GUT scale each with five texture zeroes in the combined U-D system that are consistent with the observed experimental values for the quark masses and CKM angles.

We show that if the Yukawa coupling, $\lambda_t$, of the top quark goes to infinity at the GUT scale then, in order to be symmetric at that scale, the Yukawa matrices must have texture zeroes. The U-D- system then resembles solution 4 of RRR. Specifically, we find

$$U = \begin{pmatrix} o & \alpha & o \\ \alpha & \beta & \gamma \\ o & \gamma & \delta \end{pmatrix}; \quad D = \begin{pmatrix} o & a & o \\ a & b & o \\ o & o & d \end{pmatrix} \quad (I.1)$$

where the matrix elements above are functions of t = ln (μ/1 Gev). All the texture zeroes, except for that of $D_{11}$, can be attributed to the infinite $\lambda_t$ boundary condition in the solution of the RG equation at the GUT scale.

With regard to our assumption about an infinite $\lambda_t$ at the GUT scale we note that there are physical circumstances under which such a behavior is predicted. For example, Bardeen et al [4][5] have observed that the existence of a Higgs boson as a t condensate within a SUSY-GUT model implies

$$\lambda_t \to \infty \quad as \quad \mu \to \wedge$$

where the mechanism responsible for condensate formation is assumed to arise from GUT scale dynamics. Our assumption about Yukawa matrices then implies that the dynamics that gives rise to the condensate also gives rise to symmetric Yukawa matrices at the GUT scale.

Before proceeding to discuss our model in detail we summarize below the role of texture zeroes in a 2 x 2 system and then discuss the main results of RRR.

For a 2 x 2 matrix there exists the so-called sea-saw mechanism [6] which generates the mass-hierarchy and relates the mixing angle to these masses. Here the mass matrix is symmetric and has a zero in (11)-position,

$$Y = \begin{pmatrix} o & b \\ b & d \end{pmatrix}; \quad b << d$$

If we write the diagonalized version as

$$Y^{diag} = \begin{pmatrix} m_1 & o \\ o & m_2 \end{pmatrix}$$

then

$$m_1 = b^2/d, \quad m_2 = d \tag{I.2}$$

which exhibits the desired mass hierarchy, $m_1 << m_2$. For the diagonalizing matrix

$$R = \begin{pmatrix} c & -s \\ s & c \end{pmatrix}; \quad c = \cos\theta, \quad s = \sin\theta$$

one finds

$$s \approx \frac{b}{d} \tag{I.3}$$

Expressing it in terms of $m_1$ and $m_2$ we find

$$s = \sqrt{\frac{m_1}{m_2}} \tag{I.4}$$

which relates the mixing angle to the ratio of masses.

The above example of a single texture zero in a symmetric 2 x 2 matrix illustrates the crucial role these zeroes play in relating different experimental quantities. By reducing the total number of independent parameters in Y to just two it establishes a relationship between the three parameters $m_1$, $m_2$, and s given by (I.4)

For a more general symmetric 2 x 2 matrix

$$Y = \begin{pmatrix} a & b \\ b & d \end{pmatrix} \tag{I.5}$$

with $a \ll b \ll d$, the relation (I.2) remains correct but to satisfy (I.3) and (I.4) one must have

$$a \ll b^2/d \tag{I.6}$$

in which case the situation is equivalent to having a texture zero (a = o)

For the more relevant case of a 3 x 3 matrix RRR, in the above-mentioned investigation, find five separate solutions consistent with experiments, each with a total of five texture zeroes for the combined U - D - system.

Our prediction for U and D given by (I.1) corresponds to solution 4 of RRR whose general structure, for both U and D, is with $x \ll y$, $z \ll w$

$$Y = \begin{pmatrix} o & x & o \\ x & y & z \\ o & z & w \end{pmatrix} \tag{I.7}$$

Reviewing the discussion in RRR we note that the eigenvalues $m_1$, $m_2$, and $m_3$, in

$$Y^{diag} = \begin{pmatrix} m_1 & o & o \\ o & m_2 & o \\ o & o & m_3 \end{pmatrix}$$

are given by

$$m_3 = w, \quad m_2 = y - \frac{z^2}{w}, \quad m_1 = \frac{x^2}{m_2} \tag{I.8}$$

and the diagonalizing matrix, written in the same manner as in the 2 x 2 sea-saw case, is given by

$$R = \begin{pmatrix} c_\alpha & -s_\alpha & o \\ s_\alpha & c_\alpha & o \\ o & o & 1 \end{pmatrix} \begin{pmatrix} 1 & o & o \\ o & c_\beta & -s_\beta \\ o & s_\beta & c_\beta \end{pmatrix} \tag{I.9}$$

As with the 2 x 2 case one can show that

$$s_\alpha = \frac{x}{y - \frac{z^2}{w}}$$

$$s_\beta = \frac{z}{w} \tag{I.10}$$

In the following, for the D-sector, the rotation matrix will be designated as $R_d$ and the corresponding angles as $\alpha = 1$, $\beta = 4$. For the U-sector the corresponding quantities will be taken as $R_u$ with $\alpha = 2$, $\beta = 3$. The CKM matrix can now be constructed as

$$Vckm = R_u \, P \, R_d^+ \tag{I.11}$$

where P is the (diagonal) matrix involving the phases i.e.,

$$P = \begin{pmatrix} e^{i\theta} & o & o \\ o & e^{i\theta} & o \\ o & o & 1 \end{pmatrix} \begin{pmatrix} 1 & o & o \\ o & e^{-i\phi} & o \\ o & o & e^{-i\phi} \end{pmatrix} \quad (\text{I.12})$$

The resulting CKM matrix (I.11) is

$$Vckm \approx \begin{pmatrix} c_1 c_2 - s_1 s_2 e^{-i\phi} & s_1 e^{i\phi} + c_1 s_2 & s_2(s_3 - s_4 e^{i\theta}) \\ -c_1 s_2 - s_1 e^{-i\phi} & -s_1 s_2 e^{i\phi} + c_1 c_2 c_3 c_4 + s_3 s_4 e^{-i\theta} & s_3 - s_4 e^{i\theta} \\ s_1(s_3 - s_4 e^{-i\theta}) & -c_1(s_3 - s_4 e^{-i\theta}) & c_3 c_4 + s_3 s_4 e^{i\theta} \end{pmatrix} \quad (\text{I.13})$$

## II Renormalization Group Equations for U, D and Vckm

The RG equations for U and D in the MSSM scheme are well known and are given to one-loop [2][7] by

$$16\pi^2 \frac{dU}{dt} = [-\sum_i c_i g_i^2 + 3UU^+ + DD^+ + Tr(3UU^+)] U \quad (\text{II.1})$$

$$16\pi^2 \frac{dD}{dt} = [-\sum_i c_i' g_i^2 + 3DD^+ + UU^+ + Tr(3DD^+)] D \quad (\text{II.2})$$

where we have ignored the leptonic sector. The $g_i$'s are the gauge couplings and the constants $c_i$, $c_i'$ are given by

$$c_i = \left( \frac{13}{15}, 3, \frac{16}{3} \right)$$

$$c_i' = \left( \frac{7}{15}, 3, \frac{16}{3} \right) \quad (\text{II.3})$$

Olechewski and Pokorski [8] have derived the RG equations for Vckm. Expressing the matrix elements in the form $V_{ij}$ (i,j = 1,2,3) they obtain, for MSSM

$$16\pi^2 \frac{d}{dt} |V_{ij}| = -(\lambda_t^2 + \lambda_b^2)|V_{ij}| \quad (ij = 13, 31, 23, 32) \quad (\text{II.4})$$

Of particular interest to us as we shall see later, are the ratios $|V_{31}|/|V_{23}|$ and $|V_{13}|/|V_{23}|$. From the above RG equations it can easily be shown that the t-derivatives of these ratios identically vanish. By expressing the matrix elements in terms of quark indices this implies that

$$\frac{|V_{td}|}{|V_{cb}|} = const.$$

$$\frac{|V_{ub}|}{|V_{cb}|} = const.$$

(II.5)

The RG equation for $V_{12}$ gives

$$\frac{d}{dt}|V_{12}| \sim O(V_{12}^4)$$

(II.6)

The term on the right hand side above is extremely small ($V_{12} \approx 0.22$ at low energies). If we neglect it then we obtain

$$|V_{us}| \approx const.$$

(II.7)

We turn to the quark mass matrices in the next two sections.

III  The Quark Masses at the GUT Scale

We now solve for the eigenvalues (the quark masses or the corresponding Yukawa couplings $\lambda_i$) of U and D from (II.1) and (II.2). If we take

$$\lambda_t >> \lambda_i \quad (i = u, c, d, s, b)$$

then, ignoring the gauge and D- contributions in (II.1) near the GUT scale, one can solve the RG equations for the eigenvalues of U given by

$$U^{diag} = \begin{pmatrix} \lambda_u & o & o \\ o & \lambda_c & o \\ o & o & \lambda_t \end{pmatrix}$$

For $\lambda_t$ one obtains

$$16\pi^2 \frac{d\lambda_t}{dt} = 3\lambda_t^3 + 3(\lambda_u^2 + \lambda_c^2 + \lambda_t^2)\lambda_t^2 \tag{III.1}$$

Keeping only the most dominant term, we have

$$\frac{d\lambda_t}{dt} = \frac{3\lambda_t^3}{8\pi^2} \tag{III.2}$$

If, as discussed in the Introduction, we impose the boundary condition at the GUT scale

$$\lambda_t \to \infty \quad as \quad t \to \ln \wedge \tag{III.3}$$

then

$$\lambda_t = \frac{2\pi}{\sqrt{3}} \frac{1}{(\ln\wedge - t)^{1/2}} \tag{III.4}$$

This is the same result as obtained by Bardeen et al [4][5]. We note that there are no arbitrary constants due to the nature of the non-linear equation and the boundary condition.

$$16\pi^2 \frac{d\lambda_c}{dt} = 3\lambda_c^3 + 3(\lambda_u^2 + \lambda_c^2 + \lambda_t^2)\lambda_c \tag{III.5}$$

As for the other two couplings we have for $\lambda_c$, near $\mu = \Lambda$
and similarly for $\lambda_u$. To obtain the behavior of $\lambda_c$ vis-a-vis $\lambda_t$ one notes that the gauge contributions cancel out from the RG equations for the ratio

$$R_{ct} = \frac{\lambda_c}{\lambda_t}$$

Consequently, keeping, once again, only the most dominant term one obtains from (III.1) and (III.5)

$$\frac{dR_{ct}}{dt} = \frac{-3\lambda_t^2}{16\pi^2} R_{ct} \tag{III.6}$$

which gives

$$R_{ct} = const. \, (\ln \wedge - t \,)^{1/4}$$

Thus, unlike $\lambda_t$, $\lambda_c$ is given only up to an arbitrary multiplicative constant. It is interesting to and therefore,

$$\lambda_c = \frac{const.}{(\ln \wedge - t \,)^{1/4}} \qquad (\text{III}.7)$$

note from (III.7) that

$$\lambda_c \to \infty \quad as \quad t \to \ln \wedge$$

but with a smaller power than $\lambda_t$. As noted above the gauge couplings are canceled out exactly in (III.6) so that, as one evolves from the GUT scale to lower energies, there are no gauge terms present in $R_{ct}$. The gauge contributions to $\lambda_c$ is, therefore, inherited entirely through $\lambda_t$.

Similarly for $\lambda_u$,

$$R_{ut} = const. \, (\ln \wedge - t \,)^{1/4}$$

and

$$\lambda_u = \frac{const.}{(\ln \wedge - t \,)^{1/4}} \qquad (\text{III}.8)$$

The arbitrary constants can be determined from the experimentally known values for the masses $m_c$ and $m_u$.

For the D-sector we write

$$D^{diag} = \begin{pmatrix} \lambda_d & o & o \\ o & \lambda_s & o \\ o & o & \lambda_b \end{pmatrix}$$

and use (II.2). Even though U and D cannot be diagonlized simultaneously we can, to a very good approximation, take $UU^+$ to be given entirely by $\lambda_t$ so that the only non-zero contribution from U in (II.2) comes from $(UU^+)_{33} = \lambda$.

Solving (II.2) at the GUT scale keeping only the $UU^+$ contribution, one obtains

$$\lambda_b = \frac{const.}{(\ln\wedge - t)^{1/12}} \qquad (\text{III}.9)$$

and, through the ratios $R_{cb}$ and $R_{db}$,

$$\lambda_s = const., \qquad \lambda_d = const. \qquad (\text{III}.10)$$

One can also consider the ratios $R_{uc}$ ($=\lambda_u/\lambda_c$) and $R_{ds}$ ($=\lambda_d/\lambda_s$). In the RG equations for $R_{uc}$ ($R_{ds}$) the gauge couplings as well as the dominant $\lambda_t$ and $\lambda_b$ contributions cancel out so that

$$\frac{dR_{uc}}{dt} = 0 = \frac{dR_{ds}}{dt}$$

Thus these ratios are constant consistent with the results obtained earlier. Converting them to masses we find that since the β-dependent terms divide out in each of the two ratios that

$$\frac{m_u}{m_c} = const.$$
$$\frac{m_d}{m_s} = const. \qquad (\text{III}.11)$$

### IV  The Full U- and D- Matrices at the GUT Scale and the Texture Zeroes

Consider the RG equation for the full D-matrix near the GUT scale in which only the $\lambda_t$ - contribution is kept[2][7].

We then have

$$16\pi^2 \frac{dD}{dt} = UU^+ D = \begin{pmatrix} o & o & o \\ o & o & o \\ o & o & \lambda_t^2 \end{pmatrix} D \qquad (\text{IV}.1)$$

For $D_{33}$ this gives the same results as $\lambda_b$

$$D_{33} \sim \frac{1}{(\ln\wedge - t)^{1/12}} \qquad (\text{IV}.2)$$

For $D_{31}$ and $D_{13}$ we find

$$16\pi^2 \frac{dD_{31}}{dt} = \lambda_t^2 D_{31} \quad ; \quad 16\pi^2 \frac{dD_{13}}{dt} = 0$$

Substituting the expression for $\lambda_t$ from (III.4) the solutions are given by

$$D_{31} = \frac{c_{31}}{(\ln\Lambda - t)^{1/12}} \quad , \quad D_{13} = c_{13} \tag{IV.3}$$

where $c_{31}$ and $c_{13}$ are constants.

If, as we have assumed, the Yukawa matrices are symmetric at the GUT scale then, at $t = \ln\Lambda$, one must have

$$D_{31} = D_{13} \tag{IV.4}$$

A comparison of the equations (IV.3) and (IV.4) at $t = \ln\Lambda$ clearly shows that the two can be reconciled only if $c_{13} = c_{31} = 0$ [10]. Therefore, at the GUT Scale

$$D_{31} = D_{13} = 0$$

Similarly, one can show that

$$D_{32} = D_{23} = 0$$

The remaining matrix elements of D are constants [11] which, however, can be made to satisfy the symmetry requirement without being zero.

The structure of the D-matrix is then

$$D = \begin{pmatrix} D_{11} & D_{12} & 0 \\ D_{21} & D_{22} & 0 \\ 0 & 0 & D_{33} \end{pmatrix}$$

It has two texture zeroes which, as we demonstrated, arise naturally as a combined consequence of the $\lambda_t$ - boundary condition and the symmetry requirement at the GUT Scale.

Since the sub-matrix $D_{ij}$ (i,j = 1,2) is in a (disjointed) block form and its matrix elements are constants, we can choose it to be of the sea-saw form, discussed in the Introduction, by assuming $D_{11} = 0$.

We can then write the D-matrix at the GUT Scale in the form

$$D = \begin{pmatrix} o & a & o \\ a & b & o \\ o & o & d \end{pmatrix} \qquad (IV.5)$$

where,

$$a, b \sim const. \qquad (IV.6)$$

$$d \sim \frac{1}{(\ln\Lambda - t)^{1/12}}$$

We write the U-matrix in the symmetric form at the GUT Scale

$$U = \begin{pmatrix} \eta & \alpha & \varepsilon \\ \alpha & \beta & \gamma \\ \varepsilon & \gamma & \delta \end{pmatrix} \qquad (IV.7)$$

and consider the RG equation (II.1) at the GUT Scale, ignoring the gauge and the D-contributions,

$$16\pi^2 \frac{dU}{dt} = [3UU^+ + Tr(3UU^+)]U$$

It is easy to check that the right hand side is symmetric and, as a result, so is the left hand side.

In the approximation that $\delta$ is dominant one finds

$$\delta = \frac{2\pi}{\sqrt{3}} \frac{1}{(\ln\Lambda - t)^{1/2}} \qquad (IV.8)$$

which, not surprisingly, is the same expression as that of $\lambda_t$ in (III.4).

One can also determine $\gamma$ to be

$$\gamma = \frac{a_1}{(\ln\wedge - t)^{1/2}} \tag{IV.9}$$

where $a_1$ is a constant which, in order to maintain the correct hierarchy, must satisfy $\gamma \ll \delta$, i.e.,

$$a_1 \ll \frac{2\pi}{\sqrt{3}} \tag{IV.10}$$

In Appendix A it is shown that

$$\beta = \frac{a_1^2 \sqrt{3}}{2\pi} \frac{1}{(\ln\wedge - t)^{1/2}} + \frac{a_2}{(\ln\wedge - t)^{1/4}} \tag{IV.11}$$

where $a_2$ is a constant and $a_1$ is the same constant as in (IV.9). Both the terms above must be kept in order to give the correct behavior for the eigenvalue $\lambda_c$ because, when one diagonalizes U, one obtains

$$\lambda_c = \beta - \frac{\gamma^2}{\delta}$$

$$\sim \frac{1}{(\ln\wedge - t)^{1/4}} \tag{IV.12}$$

which reproduces the behavior derived in (III.7).

In appendix A it is also shown that

$$\alpha, \varepsilon \sim \frac{1}{(\ln\wedge - t)^{1/4}}$$

$$\eta \sim const. \tag{IV.13}$$

Thus $\eta$ in comparison to the other matrix elements as well as to $\lambda_u$ (see III.8) is small at the GUT Scale and can, therefore, be neglected.

In Appendix B it is shown, furthermore, that because $\varepsilon$ is of the same order as $\alpha$ its contribution to both the mass eigenvalues and to Vckm is negligible compared to that of $\alpha$.

Thus, to an excellent approximation, one can take

$$\eta, \varepsilon \sim 0 \qquad (IV.14)$$

In summary then, the structure of the U-matrix at the GUT Scale is given by

$$U = \begin{pmatrix} o & \alpha & o \\ \alpha & \beta & \gamma \\ o & \gamma & \delta \end{pmatrix} \qquad (IV.15)$$

where

$$\alpha \sim \frac{1}{(\ln \wedge - t)^{1/4}}, \quad \beta \sim \frac{1}{(\ln \wedge - t)^{1/2}}, \quad \gamma \sim \frac{1}{(\ln \wedge - t)^{1/2}} \qquad (IV.16)$$

and

$$\delta = \frac{2\pi}{\sqrt{3}} \frac{1}{(\ln \wedge - t)^{1/2}}$$

## V. Determination of Vckm and the Predictions of the Model

As demonstrated in the previous sections our model predicts the U-D matrices to have the same structure as that of solution 4 of RRR.

One can take over the results already derived in the Introduction for this case, noting that, for Vckm, $s_1$ and $s_4$ are involved with the D-sector, and $s_2$ and $s_3$ with the U-sector.

Comparing (IV.5) with (I.7) we find $s_1$ and $s_4$ from (I.10)

$$s_1 = \sqrt{\frac{m_d}{m_s}}$$

$$s_4 = 0 \qquad (V.1)$$

Since the ratio $m_d/m_s$ is a finite constant at the GUT scale and remains so as we evolve down to the lowest energy, $\mu \sim 1$ Gev, consistent with perturbation theory, $s_1$ is also a constant which can be determined from the experimentally observed mass values.

Similarly, comparing (IV.15) with (I.7) we find from (I.10)

$$s_2 = \sqrt{\frac{m_u}{m_c}} \qquad (V.2)$$

which once again remains a constant down to $\mu \sim 1$ Gev. However, one cannot predict $s_3(=\gamma/\delta)$ from the quark masses.

With $s_4 = 0$ the CKM matrix of (I.13) reduces to

$$Vckm \approx \begin{pmatrix} c_1 c_2 - s_1 s_2 e^{-i\phi} & s_1 e^{i\phi} + c_1 s_2 & s_2 s_3 \\ -c_1 s_2 - s_1 e^{-i\phi} & -s_1 s_2 e^{i\phi} + c_1 c_2 c_3 c_4 & s_3 \\ s_1 s_3 & -c_1 s_3 & c_3 c_4 \end{pmatrix} \qquad (V.3)$$

While two of the four parameters, $s_1$ and $s_2$ are predicted in terms of the quark mass ratios, one has to determine $s_3$ and the phase angle, $\phi$, from the experimental data.

The model predictions for $s_1$ and $s_2$ can be translated into two predictions for the ratios of the CKM matrix elements i.e.,

$$\begin{aligned} \left|\frac{V_{ub}}{V_{cd}}\right| &= s_2 = \sqrt{\frac{m_u}{m_c}} \\ \left|\frac{V_{td}}{V_{cd}}\right| &= s_1 = \sqrt{\frac{m_d}{m_s}} \end{aligned} \qquad (V.4)$$

Since the ratios on the left and right sides above are scale-independent (see (II.5) and (III.11)) the above relations derived at the GUT scale remain valid at low energies and can be compared with the measured values.

The current estimates of the running quark masses at an energy scale of $\mu \sim 1$ Gev are [12]

$$\begin{aligned} m_u &= 5.1 \pm 0.9 \, MeV, & m_d &= 9.3 \pm 1.4 \, MeV \\ m_s &= 175 \pm 25 \, MeV, & m_c &= 1230 \pm 50 \, MeV \end{aligned} \qquad (V.5)$$

where the u, d masses are estimated from chiral perturbation theory [13] and the s, c masses via QCD sum rules [14]. The resulting ratios give

$$s_1 = 0.23, \quad s_2 = 0.064 \tag{V.6}$$

with an uncertainty of ± 10% in each case. These are in good agreement with the CKM matrix elements [15].

$$\left|\frac{V_{td}}{V_{cb}}\right| = 0.22 \pm 0.08, \quad \left|\frac{V_{ub}}{V_{cb}}\right| = 0.08 \pm 0.02 \tag{V.7}$$

particularly for the 2nd quantity. Here the 1st ratio has been obtained from

$$|V_{td}| = 0.009 \pm 0.003, \quad |V_{cb}| = 0.041 \pm 0.003 \tag{V.8}$$

where the estimate of $V_{td}$ is from $B_d^o - \bar{B}_d^o$ mixing assuming the dominant contribution to come from Standard Model physics (t-exchange). The error bar comes mainly from the theoretical uncertainty in the hadronic matrix elements. Note that due to the small values of $s_2$ relative to $s_1$, one can predict the magnitude of $V_{us}$ to be in the range $s_1 \pm s_2$ irrespective of the phase angle $\phi$, which is again in agreement with the experimental value

$$|V_{us}| = 0.2205 \pm 0.0018 \tag{V.9}$$

Note that the ratios of the quark masses as well as those of the CKM matrix elements of (V.4) are scale independent quantities. The same is true of $|V_{us}|$. Therefore, one can make the above comparison at any desired scale of energy. A recent compilation of the quark masses at $\mu = M_z$ [12] gives

$$m_u = 2.33^{+.42}_{-.45} \text{ MeV}, \quad m_d = 4.69^{+.60}_{-.66} \text{ MeV},$$

$$m_s = 93.4^{+11.8}_{-13.0} \text{ MeV}, \quad m_c = 677^{+56}_{-61} \text{ MeV} \tag{V.10}$$

one can easily check that the resulting $s_1$ and $s_2$ are in close agreement with (V.6) and hence with the CKM matrix element of (V.7) and (V.9).

Finally, the unknown parameters $s_3$ is fixed by the measured value of $|V_{cb}| = 0.041 \pm .003$, while the phase angle $\phi$ can be determined from

$$|V_{us}| = (s_1^2 + s_2^2 + 2 s_1 s_2 \cos\phi)^{1/2} \tag{V.10}$$

In particular the precise value of the mass ratio $m_s/m_d = 18.9 \pm 0.8$ as suggested by Leutwyler [13] implies $s_1$ to be given by (V.1) to ± 2%. This would imply

$$\cos\phi \sim -0.3 \qquad (V.11)$$

One can re-express the above quantities in terms of the Wolfenstein parameters for the CKM matrix.

$$Vckm = \begin{pmatrix} 1-\frac{\lambda^2}{2} & \lambda & A\lambda^3(\rho - i\eta) \\ -\lambda & 1-\frac{\lambda^2}{2} & A\lambda^2 \\ A\lambda^3(1-\rho - i\eta) & -A\lambda^2 & 1 \end{pmatrix}$$

Thus

$$\begin{aligned} s_1 &= \left|\frac{V_{td}}{V_{cb}}\right| = \lambda\sqrt{(1-\rho)^2 + \eta^2} \\ s_2 &= \left|\frac{V_{ub}}{V_{cb}}\right| = \lambda\sqrt{\rho^2 + \eta^2} \\ s_3 &= |V_{cb}| = A\lambda^2 \\ \cos\phi &= \frac{\lambda^2 - s_1^2 - s_2^2}{2\, s_1 s_2} \end{aligned} \qquad (V.12)$$

The measurements for the CP-violating phase are often expressed in terms of $\rho$ and $\eta$. By taking the values of $s_1$ and $s_2$ from (V.6) and of $\lambda = (|V_{us}|)$ from (V.9) we obtain $\rho$ and $\eta$ as follows

$$-.08 < \rho < .08 \; ; \qquad -.25 < \eta < .25 \qquad (V.11)$$

This compares favorable with the measured values [16]

$$\rho = .10^{+.13}_{-.39}, \qquad \eta = .34^{+.06}_{-.09} \qquad (V.12)$$

It is clear from the RG equations of the CKM parameters that only $s_1$, $s_2$ and $\phi$ (and, therefore, $\rho$ and $\eta$) are scale independent quantities.

For the sake of completeness, we shall conclude with a brief discussion of the quark masses and their scale dependence in our model.

## VI. QUARK MASSES

The top-Yukawa coupling evolution down from an infinite GUT-scale value to low energy has already been discussed by Bardeen et al [4][5]. A rough estimate of $\lambda_t$ at $\mu = 10^2$ Gev can be obtained based on quasi-IR fixed point in the RG equations for $\lambda_t$.

In the Standard Model this is given by $\lambda_t = 4/3 \, g_3$, keeping only the SU(3) gauge coupling, $g_3$. This yields $\lambda_t = 1.6$ or $m_t = \lambda_t \, v/\sqrt{2} = 278$ Gev where v(=246 Gev) is the Higgs vacuum expectation value.

In the Minimal Supersymmetric Standard Model of present interest, the fixed point value of $\lambda_t$ is given in terms of the dominant gauge coupling $g_3$ as $\lambda_t = \sqrt{8}/3 \, g_3 = 1.13$, the resulting top quark mass is

$$m_t = \frac{\lambda_t \, v \sin \beta}{\sqrt{2}} = 197 \sin \beta \ \ GeV \tag{VI.1}$$

where $\tan \beta$ is the ratio of the vacuum-expectation values of the two Higgs doublets of MSSM. This is consistent with the experimental value of $m_t = 175$ GeV.

As far as the remaining quark-Yukawa couplings are concerned, each has an arbitrary multiplicative constant which can be determined by fitting them to the experimentally measured values. However, $m_d/m_s$ and $m_u/m_c$ remain a constant as they evolve down from the GUT Scale. It is interesting to note that $\lambda_c$, $\lambda_u$ and $\lambda_b$ (though not $\lambda_s$ or $\lambda_d$) are divergent at $\mu = \Lambda$ leading to the question whether cc, uu and bb - condensates exist or, since they are coupled through unitarity to t_, whether they contribute to the same (t_) condensate.

## VII. CONCLUSION

We have shown, using the one loop MSSM renormalization group equations, that the combined effect of a symmetric Yukawa matrix and an infinite $\lambda_t$ at the GUT Scale produces texture zeroes in U and D through a set of hierarchical solutions. The U-D system then resembles one of the solutions (solution 4) analyzed by RRR. The experimentally measured scale-independent ratios $|V_{td}|/|V_{cb}|$ and $|V_{ub}|/|V_{cb}|$ show excellent agreement with their predicted values $\sqrt{m_d/m_s}$ and $\sqrt{m_u/m_c}$ respectively. Other predictions of the model compare favorably with experiments.

## VIII. ACKNOWLEDGEMENT

One of us (BRD) thanks Dr. E. Keith and Dr. A. Khachatourian for discussions. We also thank Dr. E. Ma for his comments. This work was supported in part by the U.S. Department of Energy under grant number DE-FG0394ER40837.

## Appendix A

We first note that the solution of the equation

$$16\pi^2 \frac{d\lambda}{dt} = 3N\delta^2 \lambda, \quad \text{(A.1)}$$

where N is a constant (=1 or 2 in our case) and $\delta$ the (33) element of U,

$$\delta = \frac{2\pi}{\sqrt{3}} \frac{1}{(\ln \Lambda - t)^{1/2}} \quad \text{(A.2)}$$

is given by

$$\lambda = \frac{const.}{(\ln \Lambda - t)^{N/4}} \quad \text{(A.3)}$$

The equations (II.1) satisfied by the individual matrix elements of U in (IV.7) from (IV.12), keeping only the $\delta$ and $\delta^2$ terms, (neglecting gauge and D-contributions) are

$$16\pi^2 \frac{d\beta}{dt} = 3\gamma^2 \delta + 3\beta \delta^2$$

$$16\pi^2 \frac{d\alpha}{dt} = 3\varepsilon\gamma\delta + 3\alpha \delta^2$$

$$16\pi^2 \frac{d\eta}{dt} = 3\varepsilon^2 \delta + 3\eta \delta^2 \quad \text{(A.4)}$$

$$16\pi^2 \frac{d\varepsilon}{dt} = 3\alpha\gamma\delta + 6\varepsilon \delta^2$$

$$16\pi^2 \frac{d\gamma}{dt} = 3(\alpha\varepsilon + \beta\gamma)\delta + 6\gamma \delta^2$$

A priori if we keep only the $\delta^2$ -term above then from (A.1) and (A.3) we obtain

$$\beta \sim \alpha \sim \eta \sim \frac{1}{(\ln \Lambda - t)^{1/4}}; \quad \varepsilon \sim \gamma \sim \frac{1}{(\ln \Lambda - t)^{1/2}} \quad (A.5)$$

However, a closer look at the driving terms in the equations above (e.g. the first term proportional to δ) shows that they can alter the behavior from what is given by (A.5)

We show in what follows that there exists a set of self-consistent hierarchical solutions in which $\gamma \gg \varepsilon$ imply $\beta \gg \alpha$ and vice versa, and that it is these solutions that are relevant to our problem.

In particular we will show that

$$\eta \sim const.; \quad \alpha, \varepsilon \sim \frac{1}{(\ln \Lambda - t)^{1/4}}; \quad \beta \sim \frac{1}{(\ln \Lambda - t)^{1/2}}$$

as well as

$$\gamma \sim \frac{1}{(\ln \Lambda - t)^{1/2}}$$

(i) β and γ

In equation (A.4) if we neglected the driving term in equation for γ, we obtain from (A.1) and (A.3)

$$\gamma = \frac{a_1}{(\ln \Lambda - t)^{1/2}} \quad (A.6)$$

we will come back later to the question of the driving term.

For β, if we keep the driving term then the equation for it is of the form

$$\frac{d\lambda}{dt} = A + \lambda B \quad (A.7)$$

whose solution is

$$\lambda = [\int^t dt'\, A e^{-\int^{t'} dt''\, B} + A']\, e^{\int^t dt'\, B} \tag{A.8}$$

where $A'$ is a constant, and $A$ and $B$ are functions of t.

The solution for $\beta$, after substituting the expression for $\gamma$ and $\delta$ and using (A.8) is,

$$\beta = \frac{a_1^2 \sqrt{3}}{2\pi} \frac{1}{(\ln \Lambda - t)^{1/2}} + \frac{a_2}{(\ln \Lambda - t)^{1/4}} \tag{A.9}$$

where $a_2$ is a constant and $a_1$ is the same constant which appears in (A.6)

(ii) $\underline{\alpha,\, \varepsilon,\, \eta}$

In the equation for $\alpha$ in (A.4) if we neglect the driving term, we obtain

$$\alpha \sim \frac{1}{(\ln \Lambda - t)^{1/4}} \tag{A.10}$$

we will come back later to the assumption about the driving term.

To consider the behavior of $\varepsilon$ at the GUT scale we first note that if we ignored the driving term then we would get the behavior given by (A.5)

From (A.4) we combine the equations for $\gamma$ and $\varepsilon$ (keeping their driving terms) to write the equation for their ratio,

$$x = \frac{\varepsilon}{\gamma} \tag{A.11}$$

which is given by

$$16\pi^2 \frac{dx}{dt} = 3\alpha\delta (1 - x^2) - 3\beta\delta\, x \tag{A.12}$$

If $\varepsilon$ were of the type given by (A.5) and $\gamma$ given by (A.6) then keeping the leading $\left(\approx \dfrac{1}{(\ln \Lambda - t)^{1/2}}\right)$ as well as next to the leading $\left(\approx \dfrac{1}{(\ln \Lambda - t)^{1/4}}\right)$ terms in both we would expect the ratio x to behave as

$$x = a + b(\ln \Lambda - t)^{\frac{1}{4}} \tag{A.13}$$

However, then the left hand side (LHS) of (A.12) would behave as

$$LHS \approx \frac{1}{(\ln \Lambda - t)^{3/4}} \tag{A.14}$$

and the right hand side (RHS) from the knowledge of $\alpha$, $\beta$ and $\delta$, as

$$RHS \approx \frac{const.(1-a^2)}{(\ln \Lambda - t)^{3/4}} - \frac{const.\, a}{(\ln \Lambda - t)} \tag{A.15}$$

To satisfy equation (A.12) (LHS=RHS), one must have, a=0, and, therefore,

$$x = b(\ln \Lambda - t)^{1/4} \tag{A.16}$$

Given the behavior of $\gamma$ already determined in (A.6), this implies that

$$\varepsilon \sim \frac{1}{(\ln \Lambda - t)^{1/4}} \tag{A.17}$$

so that $\varepsilon$ does not have the behavior of the type given by (A.5) but rather a much milder behavior given by (A.17).

To determine $\eta$, we combine the equations for $\eta$ and $\alpha$ in (A.4) and write the equation for their ratio (keeping the driving terms)

$$y = \frac{\eta}{\alpha}$$

which is of the form

$$16\pi^2 \frac{dy}{dt} = \frac{3\varepsilon^2 \delta}{\alpha} - \frac{3\varepsilon \gamma \delta}{\alpha} y \tag{A.18}$$

If the behavior for $\eta$ given by (A.5) were correct then since $\alpha$ is determined by (A.10) the ratio is given by

$$y = a' + b'(\ln \Lambda - t)^{1/4} \qquad (A.19)$$

where we kept the leading $\left(\dfrac{1}{(\ln \Lambda - t)^{1/4}}\right)$ as well as next-to-leading (~constant) terms in $\eta$ and $\alpha$.

The left-hand-side of (A.18) now behaves as

$$LHS \approx \frac{1}{(\ln \Lambda - t)^{3/4}}$$

on the other hand

$$RHS \approx \frac{1}{(\ln \Lambda - t)^{3/4}} - \frac{1}{(\ln \Lambda - t)} a'$$

therefore, to have

$$LHS = RHS$$

one must have a'=0 and therefore

$$y = b'(\ln \Lambda - t)^{1/4} \qquad (A.20)$$

so that $\eta$ does <u>not</u> have the behavior given by (A.5) but rather

$$\eta \approx const$$

at the GUT scale.

Looking <u>a posteriori</u> at assumptions made in the $\alpha$ and $\gamma$ equations of neglecting the driving terms we note, after solving the equations exactly (using (A.7) and (A.8)), that the only modifications occurring in the solutions are of the type $\ln(\ln \Lambda - t)$ which is extremely mild and can be neglected.

Finally, looking at the entire set of equations contained in (A.4) as an ensemble our conclusion about the hierarchy of the matrix elements can be made highly plausible from the nature of the hierarchy of the driving terms: $\gamma \gg \varepsilon$ in the driving terms in the first three equations of (A.4) implies $\beta \gg \alpha \gg \eta$ and, in turn, $\beta \gg \alpha$ in the driving term in the last two equations implies $\gamma \gg \varepsilon$.

## Appendix B

We show below that, for the U-matrix, even though we have

$$\alpha \sim \varepsilon \sim \frac{1}{(\ln \Lambda - t)^{1/4}}$$

the contribution of $\varepsilon$ compared to $\alpha$ is negligible in obtaining both the mass-eigenvalues of U and the relevant CKM elements(e.g. $|V_{ub}|$).

As we have already demonstrated, we can take $\eta = 0$, in the U-matrix. To facilitate comparison between the contribution of $\alpha$ and $\varepsilon$ consider two separate possibilities

$$(a)\ U = \begin{pmatrix} o & \alpha & o \\ \alpha & \beta & \gamma \\ o & \gamma & \delta \end{pmatrix}, \quad (b)\ U = \begin{pmatrix} o & o & \varepsilon \\ o & \beta & \gamma \\ \varepsilon & \gamma & \delta \end{pmatrix}$$

The rotation parameters $s_1$ and $s_4$ in CKM are involved only with the D-sector and we have already shown that

$$s_4 = 0$$

The rotation parameter $s_2$ and $s_3$, on the other hand, involve only the U-sector and the dependence of CKM parameters on those angles is different depending on whether one considers case (a) or case (b).

We will compare case (a) and case (b) in terms of contribution of $\alpha$ and $\varepsilon$ to the two physical parameters, $m_u$ and $|V_{ub}|$.

Case (a). We have already considered this case in the Introduction where we also derived the CKM matrix,(I.13). With $s_4=0$, we have from the results of the Introduction

$$(i)\ |V_{ub}| = s_2\, s_3 = \sqrt{\frac{m_u}{m_c}}\, \frac{\gamma}{\delta} \rightarrow const. \quad (B.1)$$

(ii) The contribution to $m_u$ is

$$m_u = \frac{\alpha^2}{m_c} \rightarrow \frac{1}{(\ln \Lambda - t)^{1/4}}$$

Case (b) The CKM matrix (with $s_4=0$) is

$$Vckm \approx \begin{pmatrix} c_1 c_2 & s_1 e^{i\phi} & c_3 c_4 s_2 \\ -s_1 e^{-i\phi} & -c_1 c_3 c_4 & s_3 \\ -c_1 s_2 + s_1 s_3 \varepsilon^{-i\phi} & -c_1 s_3 & c_2 c_3 c_4 \end{pmatrix} \quad (B.2)$$

where one can show that

$$s_2 = \frac{\varepsilon}{\delta}, \quad s_3 = \frac{\gamma}{\delta}$$

To compare the above quantities we find

$$(i) \quad |V_{ub}| = s_2 = \frac{\varepsilon}{\delta} \rightarrow 0$$

(ii) contribution to $m_u$ is

$$m_u = \frac{\varepsilon^2}{\delta - \frac{\gamma^2}{\beta}} = \left(\frac{\varepsilon^2}{m_c}\right)\left(\frac{\beta}{\delta}\right) \quad (B.3)$$

so that even if $\varepsilon \approx \alpha$ we find that for this case there is an extra multiplicative factor (see IV.10)

$$\frac{\beta}{\delta} \rightarrow \frac{3 a_1^2}{(2\pi)^2} << 1 \quad (B.4)$$

Therefore for both the parameters, $|V_{ub}|$ and $m_u$, the contribution coming from (b) is negligible compared to case (a)[17].

We can, therefore, for all practical purposes, take

$$\varepsilon \approx 0 \quad (B.5)$$